\documentclass[12pt,a4paper]{article}
\usepackage{amsmath}
\usepackage{bm}
\usepackage{amsfonts}
\usepackage{graphicx}
\usepackage[normal]{caption2}
\usepackage{subfigure}
\usepackage{rotating}
\usepackage{citesort}
\usepackage{lscape}
\usepackage{section}
\begin{document}
\renewcommand\arraystretch{1.1}
\setlength{\abovecaptionskip}{0.1cm}
\setlength{\belowcaptionskip}{0.5cm}
\title {Isospin effects on the energy of vanishing flow in heavy-ion collisions}
\author {Sakshi Gautam$^{1}$, Rajiv Chugh$^{1}$, Aman D. Sood$^{2}$, Rajeev K.
Puri$^{1}$,\\
Ch. Hartnack$^{2}$ and J. AIchelin$^{2}$\\
\it $^{1}$Department of Physics, Panjab University, Chandigarh
-160 014, India.\\
$^{2}$SUBATECH, \\Laboratoire de Physique Subatomique et des
Technologies Associ\'{e}es\\
Universit\'{e} de Nantes - IN2P3/CNRS - EMN \\
4 rue Alfred Kastler, F-44072 Nantes, France.\\} \maketitle
\begin{abstract}
Using the isospin-dependent quantum molecular dynamics model we
study the isospin effects on the disappearance of flow for the
reactions of $^{58}Ni$+$^{58}Ni$ and $^{58}Fe$+$^{58}Fe$ as a
function of impact parameter. We found good agreement between our
calculations and experimentally measured energy of vanishing flow
at all colliding geometries. Our calculations reproduce the
experimental data within 5\%(10\%) at central (peripheral)
geometries.
\end{abstract}
 Electronic address:~rkpuri@pu.ac.in
\newpage
\section{Introduction}
The collective transverse in-plane flow \cite{sch,gust,doss86} has
been used extensively over the past three decades to study the
properties of hot and dense nuclear matter, i.e. the nuclear
matter equation of state (EOS) as well as the in-medium
nucleon-nucleon (nn) cross section. This has been reported to be
highly sensitive toward the above-mentioned properties as well as
toward entrance channel parameters such as combined mass of the
system \cite{ogli89,blat91,andro03}, colliding geometries
\cite{pan93,rami94,bert87,luka05,zhang06} as well as incident
energy of the projectile
\cite{zhang06,wmz90,beav92,luka08,hong02}. The dependence of the
collective flow on the above-mentioned parameters has revealed
much interesting physics, especially the beam energy dependence
which has also led to its disappearance. At lower incident
energies, the dominance of the attractive mean field prompts the
scattering of the particles into negative deflection angles thus
producing negative flow whereas frequent nn collisions and
repulsive mean field at higher incident energies result in the
emission of particles into positive deflection angles and hence
yield positive flow. While going through the incident energies,
collective transverse in-plane flow disappears at a particular
incident energy termed as the \textit{balance energy} ($E_{bal}$)
or \textit{energy of vanishing flow} (EVF) \cite{krof89}. The EVF
has been studied experimentally as well as theoretically over a
wide range of mass ranging from $^{12}C$ + $^{12}C$ to $^{238}U$ +
$^{238}U$ at different colliding geometries and found to vary
strongly as a function of the combined mass of the system
\cite{mota92,west93,zhou94,zhou94a,mag00,mag100,sood} as well as a
function of the impact parameter
\cite{luka05,sull90,buta95,pak96,chugh09}.
\par
 At the same time the isospin degree of freedom plays an important role in heavy-ion
collisions (HIC) through both nn collisions and the EOS. The later
quantity is important because of its profound implications for
studying  the structure and evolution of many astrophysical
objects such as neutron stars, supernova explosions, etc. To
access the EOS as well as its isospin dependence, it is important
to describe the observables of HIC such as collective transverse
in-plane flow as well as its disappearance both of which in fact
have been found to depend on the isospin degree of freedom. The
first study showing the isospin effects on the collective flow and
EVF was reported by Li $et~ al$ \cite{li96}, where a strong
dependence of both the above-mentioned quantities was shown. The
effect was found to be more pronounced at peripheral collisions.
Later on, Pak $et ~ al$ \cite{pak97} demonstrated experimentally
the isospin effects on the collective flow and EVF at central and
peripheral colliding geometries. The theoretical calculations
using the isospin-dependent Boltzmann-Uehling-Uhlenbeck (IBUU)
model were confronted with the data. The calculations
under-predicted the experimentally measured EVF. Chen $et ~ al$
\cite{liew98} studied the effect of the isospin degree of freedom
on the collective flow and EVF using the isospin-dependent quantum
molecular dynamics (IQMD) model, which was an improved version of
the original QMD model \cite{aichqmd,qmd2}. The calculated results
were found to differ from the data at all colliding geometries.
The reason for a large deviation was attributed to the low
saturation density in the initialized nuclei which was about 0.12
$fm^{-3}$ (0.75 $\rho_{0}$) as compared to the normal saturation
density of 0.16 $fm^{-3}$ and to the fact that the mean field due
to the isospin-independent part of EOS would be more attractive at
low density. However, it has been shown in \cite{khoa92} that the
mean field potential is rather the same both at $\rho/\rho_{0}$ =
1 and 0.75 for the equations of state used in \cite{liew98}. Only
at values larger than $\rho/\rho_{0}$, the mean field potential
begins to differ. Moreover in \cite{hart98} also, it has been
shown that significant differences in the collective flow values
due to the different initial densities occur only at high incident
energies. These differences vanish in the EVF domain. Scalone
\emph{et al} \cite{scal} also studied the isospin effects on the
collective flow. Their calculations indicate towards different
neutron and proton flows (see \cite {bara} also). Their results of
EVF were in good agreement with the data at $b/b_{max}$ = 0.45. In
this paper, we reproduce for the first time all the measured EVF
for $^{58}Ni$ + $^{58}Ni$ and $^{58}Fe$ + $^{58}Fe$ systems (used
to demonstrate isospin effect in \cite{pak97} and \cite{liew98})
and also explain in part why the calculations of \cite{liew98}
using the IQMD model show a large deviation from the measured EVF
at all colliding geometries. For the present study, we use the
IQMD model \cite{hart98}.

 The IQMD model is an extension of the QMD model \cite{aichqmd,qmd2}, which treats different charge states of
nucleons, deltas and pions explicitly, as inherited from the
Vlasov-Uehling-Uhlenbeck (VUU) model \cite{vuu}. The IQMD model
has been used successfully for the analysis of a large number of
observables from low to relativistic energies. The isospin degree
of freedom enters into the calculations via symmetry potential,
cross sections and Coulomb interaction.
 \par
 In this model, baryons are represented by Gaussian-shaped density distributions
  \begin {eqnarray}
  f_{i}(\vec{r},\vec{p},t) =
  \frac{1}{\pi^{2}\hbar^{2}}\exp(-[\vec{r}-\vec{r_{i}}(t)]^{2}\frac{1}{2L})
   \times \exp(-[\vec{p}- \vec{p_{i}}(t)]^{2}\frac{2L}{\hbar^{2}})
 \end {eqnarray}
 Nucleons are initialized in a sphere with radius R = 1.12 A$^{1/3}$ fm, in accordance with the liquid-drop model.
 Each nucleon occupies a volume of \emph{h$^{3}$}, so that phase space is uniformly filled.
 The initial momenta are randomly chosen between 0 and Fermi momentum ($\vec{p}$$_{F}$).
 The nucleons of the target and projectile interact by two- and three-body Skyrme forces, Yukawa potential, Coulomb interactions and momentum
 dependent interactions. In addition to the use of explicit charge states of all baryons and mesons, a symmetry potential between protons and neutrons
 corresponding to the Bethe-Weizsacker mass formula has been included.
 The hadrons propagate using the Hamilton equations of motion:
 \begin {eqnarray}
 \frac{d\vec{{r_{i}}}}{dt} = \frac{d\langle H
 \rangle}{d\vec{p_{i}}}; ~~~~~~~~~ \frac{d\vec{p_{i}}}{dt} = - \frac{d\langle H
  \rangle}{d\vec{r_{i}}}
 \end {eqnarray}
  with
\begin{eqnarray}
\nonumber\langle H\rangle & = & \langle T\rangle
+ \langle V \rangle \\
\nonumber                 & = & \sum_{i}\frac{p^{2}_{i}}{2m_{i}} +
                   \sum_{i}\sum_{j>i}\int
                   f_{i}(\vec{r},\vec{p},t)V^{ij}
                   (\vec{r}~',\vec{r}) \\
                          &   & \times
                   f_{j}(\vec{r}~',\vec{p}~',t)
                   d\vec{r}~ d\vec{r}~'~ d\vec{p}~ d\vec{p}~'.
 \end{eqnarray}
 The baryon potential\emph{ V$^{ij}$}, in the above relation, reads as
\begin{eqnarray}
\nonumber V^{ij}(\vec{r}~'-\vec{r})& = & V^{ij}_{Skyrme} + V^{ij}_{Yukawa} +V^{ij}_{Coul} + V^{ij}_{mdi} + V^{ij}_{sym} \\
\nonumber                          & = & [t_{1}\delta(\vec{r}~'-\vec{r})+t_{2}\delta(\vec{r}~'-\vec{r})\rho^{\gamma-1}(\frac{\vec{r}~'+\vec{r}}{2})]\\
\nonumber                          &   & +t_{3}\frac{\exp(|(\vec{r}~'-\vec{r})|/\mu)}{(|(\vec{r}~'-\vec{r})|/\mu)}+\frac{Z_{i}Z_{j}e^{2}}{|(\vec{r}~'-\vec{r})|}\\
\nonumber                          &   & +t_{4}\ln^{2}[t_{5}(\vec{p}~'-\vec{p})^{2} +1]\delta(\vec{r}~'-\vec{r})\\
                                   &   & +t_{6}\frac{1}{\varrho_{0}}T_{3i}T_{3j}\delta(\vec{r_{i}}~'-\vec{r_{j}}).
\end{eqnarray}
Here t$_{6}$ = 4C with C = 32 MeV and \emph{Z$_{i}$} and
\emph{Z$_{j}$} denote the charges of the \emph{ith} and \emph{jth}
baryon, and \emph{T$_{3i}$} and
 \emph{T$_{3j}$}
 are their respective \emph{T$_{3}$} components (i.e. $1/2$ for protons and $-1/2$ for neutrons).
The parameters\emph{ $\mu$} and \emph{t$_{1}$,....,t$_{6}$} are
adjusted to the real part of the nucleonic optical potential.
 For the density dependence of  the nucleon optical potential, standard Skyrme-type parametrization is employed.
 The momentum dependence \emph{V$_{mdi}^{ij}$} of the nn interactions, which may optionally be used in IQMD, is fitted to the experimental data
 in the real part of the nucleon optical potential.
  We also use the standard energy-dependent free nn cross
  section $\sigma_{nn}^{free}$ as well as the cross section reduced by
  20$\%$, i.e. $\sigma$ = 0.8 $\sigma_{nn}^{free}$.
The details about the elastic and inelastic cross sections for
proton-proton and proton-neutron collisions can be found in
\cite{hart98,cug}. The cross sections for neutron-neutron
collisions are assumed to be equal to the proton-proton cross
sections. Two particles collide if their minimum distance\emph{ d}
fulfills
\begin {equation}
 d \leq d_{0} = \sqrt{\frac{\sigma_{tot}}{\pi}},   \sigma_{tot} =
 \sigma(\sqrt{s}, type),
\end {equation}
\begin{figure}[!t] \centering
 \vskip 1cm
\includegraphics[angle=0,width=8cm]{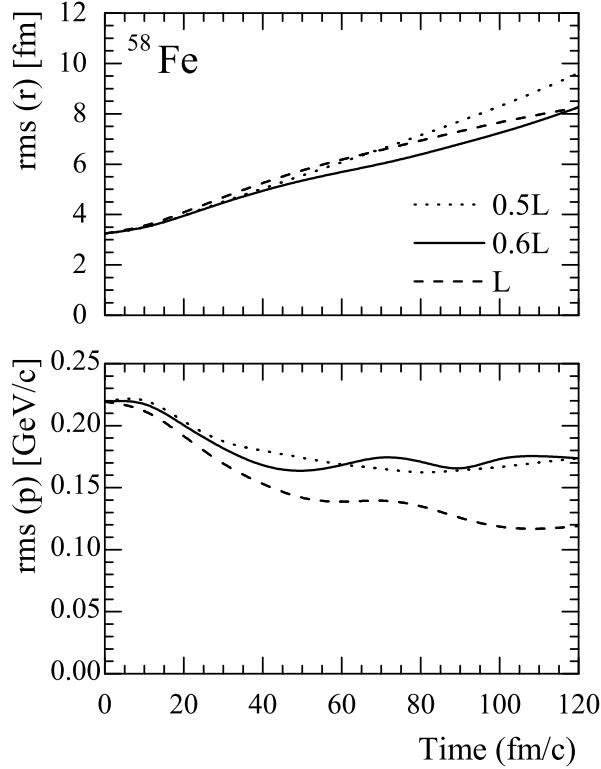}
 \vskip -0cm \caption{ Time evolution of root mean square radii
 of a single $^{58}$Fe nucleus in coordinate and momentum space obtained
 with IQMD for EOS used in the present study for different
 Gaussian widths of 0.5L, 0.6L and L. Various lines are explained in the text.
}\label{fig1}
\end{figure}
where 'type' denotes the ingoing collision partners (N-N....).
Explicit Pauli blocking is also included; i.e. Pauli blocking of
the neutrons and protons is treated separately. We assume that
each nucleon occupies a sphere in coordinate and momentum space.
This trick yields the same Pauli blocking ratio as an exact
calculation of the overlap of the Gaussians will yield. We
calculate the fractions P$_{1}$ and P$_{2}$ of final phase space
for each of the two scattering partners that are already occupied
by other nucleons with the same isospin as that of scattered ones.
The collision is blocked with the probability
\begin {equation}
 P_{block} = 1-[1 - min(P_{1},1)][1 - min(P_{2},1)],
\end {equation}
and, correspondingly is allowed with the probability 1 -
P$_{block}$. For a nucleus in its ground state, we obtain an
averaged blocking probability $\langle P_{block}\rangle$ = 0.96.
Whenever an attempted collision is blocked, the scattering
partners maintain the original momenta prior to scattering. As
mentioned in  \cite{hart98}, in IQMD the Gaussian width L (which
can be regarded as a description of the interaction range of the
particle) depends on the system under study.  The system
dependence of L in IQMD has been introduced in order to obtain the
maximum stability of the nucleonic density profile. For example,
for Au+Au (Ca+Ca and lighter nuclei) L = 2.16 (1.08) $fm^{2}$.
Therefore, in the present study we use the Gaussian width 0.6L. In
figure 1 we display the time evolution of the rms radius of a
single $^{58}$Fe nucleus in coordinate and momentum space for
different choices of L. The dotted, solid and dashed lines are for
the Gaussian width 0.5L, 0.6L and L, respectively. From the
figure, we see that $^{58}$Fe shows the maximum stability for the
Gaussian width 0.6L and is least stable for 0.5L. We also find
that the stability of a single $^{58}$Fe nucleus is quite the same
for 0.6L and L. We find similar results for the $^{58}$Ni nucleus
also (not shown here). We will come back to this point later. It
is worth mentioning that the appropriate choice of the Gaussian
width (interaction range) is very important since a choice of a
different interaction range causes different density profiles of
the ground-state nucleus which results in the different strengths
of density gradient that in turn has strong influence on the
variables such as flow, multifragmentation, pion, kaon production
etc \cite{hart98,klak93,kaon1}.
\par
We simulate 2500 events for the reactions $^{58}Ni$ + $^{58}Ni$
and $^{58}Fe$ + $^{58}Fe$ between incident energy range from 50 to
150 MeV/nucleon in small steps of 10 MeV/nucleon. The impact
parameters are guided by \cite{pak97}. We use the soft EOS along
with momentum-dependent interactions (MDI) labeled as SMD. The
reactions are followed till the transverse flow saturates. The
saturation time is around 100 fm/c for the reactions in the
present study.
 There are several methods in the literature to define the nuclear transverse in-plane flow. In most of the
 studies, the EVF is extracted from (p$_{x}$/A) plots where one plots (p$_{x}$/A) as a function of Y$_{c.m.}$/Y$_{beam}$. Using the linear fit to the slope, one can find the so-called
 reduced flow F. Naturally, the energy at which the reduced flow passes through zero is called the EVF.
 Alternatively, one can also use a more integrated quantity "\textit{directed transverse momentum $\langle
p_{x}^{dir}\rangle$}" which is defined as
\cite{sood,aichqmd,hart98,lehm}
\begin {equation}
\langle{p_{x}^{dir}}\rangle = \frac{1} {A}\sum_{i=1}^{A}{sign\{
{y(i)}\} p_{x}(i)},
\end {equation}
where $y(i)$ and $p_{x}$(i) are, respectively, the rapidity and
the momentum of the $i^{th}$ particle. The rapidity is defined as
\begin {equation}
Y(i)= \frac{1}{2}\ln\frac{{\vec{E}}(i)+{\vec{p}}_{z}(i)}
{{\vec{E}}(i)-{\vec{p}}_{z}(i)},
\end {equation}
where $\vec{E}(i)$ and $\vec{p_{z}}(i)$ are, respectively, the
energy and longitudinal momentum of the $i^{th}$ particle. In this
definition, all the rapidity bins are taken into account. It,
therefore, presents an easier way to measure the in-plane flow
rather than complicated functions such as (p$_{x}$/A) plots. It
has been shown in \cite{sood} that the disappearance of flow
occurs at the same incident energy in both the cases showing the
equivalence between p$_{x}$/A and $\langle{p_{x}^{dir}}\rangle$ as
far as the EVF is concerned. It is worth mentioning that the EVF
has the same value for all fragments types
\cite{west93,pak96,pak97,west98,cuss}. Further the apparatus
corrections and acceptance do not play any role in calculation of
the EVF \cite{ogli89,west93,cuss}.
\begin{figure}[!t]
\centering
 \vskip 1cm
\includegraphics[angle=0,width=12cm]{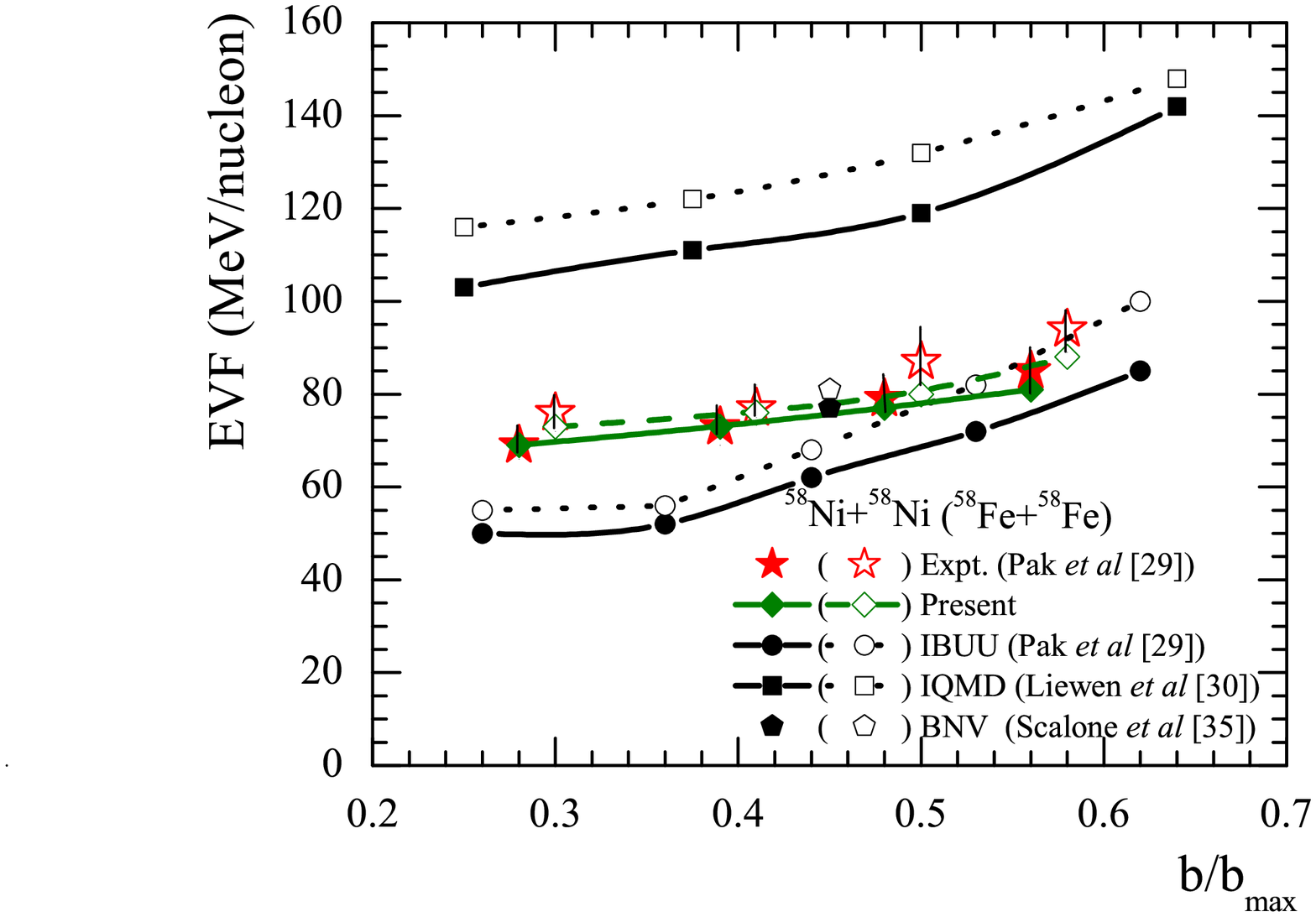}
 \vskip -0cm \caption{ The EVF as a function
of the reduced impact parameter. Various symbols are explained in
the text}\label{fig2}
\end{figure}
\par
In figure 2, we display the EVF as a function of the reduced
impact parameter $b/b_{max}$ for the reactions $^{58}Ni$ +
$^{58}Ni$ (solid symbols) and $^{58}Fe$ + $^{58}Fe$ (open
symbols). Stars represent the experimental data whereas diamonds
correspond to our theoretical calculations. Squares (circles)
represent the IQMD (IBUU) calculations of \cite{liew98}
(\cite{pak97}). Pentagons represent the theoretical calculations
of \cite{scal} for b/b$_{max}$ = 0.45. The lines are only to guide
the eye. Our results of the EVF and experimental data for the
reaction $^{58}Fe$ + $^{58}Fe$ have been slightly offset in the
horizontal direction for clarity. The vertical lines on the data
points represent statistical errors. The statistical error bars on
theoretical points of \cite{pak97} and \cite{liew98} are not
displayed, again for clarity. For the calculations of EVF, we use
the standard energy-dependent free nn cross section as was done in
\cite {pak97} and \cite {liew98} also. Our EVF values for
$\sigma_{nn}^{free}$ (not shown here) are lower than the data
consistently by about 25\%, at all colliding geometries.
Therefore, we reduce the cross section by 20\% with $\sigma$ =
0.8$\sigma_{nn}^{free}$. We find that the EVF increases with
decrease in cross section at all colliding geometries in agreement
with \cite{sood} where Sood and Puri have decomposed the total
transverse flow into contribution due to mean field and two-body
collision parts and showed that the EVF is a result of
counterbalancing of the flow due to mean field and collisions.
With a decrease in cross section, the flow due to collisions
decreases; therefore, higher incident energy is needed to
compensate this effect. From the figure we see that our
calculations are in good agreement with the data and the
calculations of \cite{scal}. Note that we have also used symmetry
energy linear in its density dependence as was done in \cite
{pak97} and \cite {liew98}.
\begin{figure}[!t]
\centering \vskip 1cm
\includegraphics[angle=0,width=10cm]{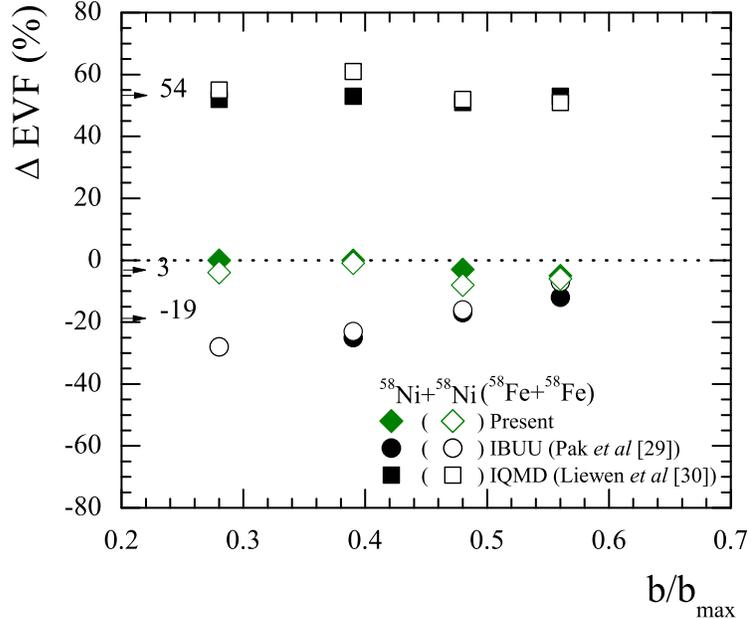}
\vskip -0cm \caption{The percentage deviation of EVF values for
different calculations over the experimentally measured EVF as a
function of the reduced impact parameter.}\label{fig3}
\end{figure}
The results of IBUU calculations \cite{pak97} under-predict the
data whereas the IQMD calculations of \cite{liew98} over-predict
the data consistently. It is worth mentioning here that the choice
of reduced cross section has also been motivated by \cite{daff} as
well as many previous studies
\cite{andro03,hong02,zhou94a,peter,akira}. In \cite {daff}, Daffin
and Bauer have suggested the factor of 0.2-0.3 for the
density-dependent reduction of the in-medium cross section. Their
theoretical results (not shown here) were much closer to the data
when using the in-medium reduction of the scattering cross
section. However the difference between EVF of $^{58}Fe$ +
$^{58}Fe$ and $^{58}Ni$ + $^{58}Ni$, at all colliding geometries,
was reduced by a factor of 4 than what was observed in
experiments. Further, we have calculated the EVF at central and
peripheral colliding geometries with the isotropic cross section.
We find the effect of an angular distribution of the scattering
cross sections on the EVF to be negligible for both $^{58}Fe$ +
$^{58}Fe$ and $^{58}Ni$ + $^{58}Ni$. It is worth mentioning here
that the effect of an angular distribution on transverse flow is
significant at higher energies \cite{hartth}. We have also
calculated the EVF for neutrons and protons for the more
neutron-rich system where a larger difference between neutron and
proton flow is expected \cite{scal}. We find that the EVF is the
same for neutrons, protons and all nucleons at central collisions.
At peripheral collisions, neutron and proton EVF differ each other
by 3 MeV and the EVF for all nucleons lie in between the neutron
and proton EVF. The results are in agreement with \cite{scal}.
Further, we have checked the effect of different symmetry energy
(by varying both the strength of symmetry energy as well as its
density dependence) on the transverse flow (due to neutrons,
protons and all nucleons) at high densities. At 150 MeV/nucleon,
although we obtain different neutron and proton flow in agreement
with \cite{scal}, the difference between neutron and proton flow
is insensitive to the choice of symmetry energy for the systems in
the present study. However, we do not exclude the possibility of
this effect for systems having a large N/Z ratio. At 400
MeV/nucleon the transverse flow is insensitive to the symmetry
energy. In the present study the effect of n/p effective mass
splitting is expected to be negligible since N/Z ratio for the two
systems in the present study is small. The above discussion
indicates that the studies with systems having a large difference
between the N/Z ratio in the Fermi energy domain could be best
suitable to explore the isospin effects of in-medium nuclear
interactions in transverse flow.
\begin{figure}[!t]
\centering \vskip 1cm
\includegraphics[angle=0,width=12cm]{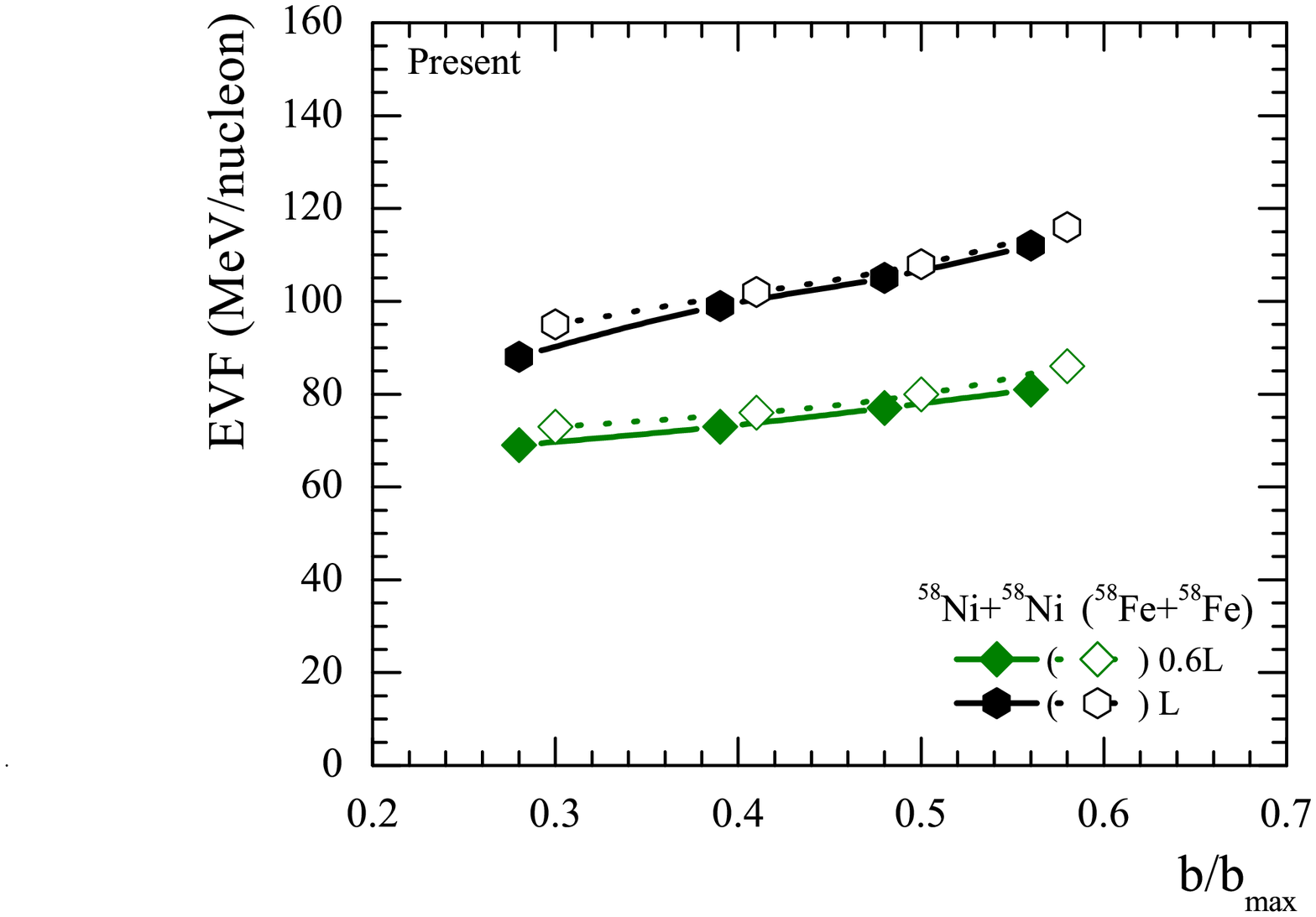}
\vskip -0cm \caption{The EVF as a function of the reduced impact
parameter for different values of L. The solid (open) symbols
represent the reactions $^{58}Ni$ + $^{58}Ni$ ($^{58}Fe$ +
$^{58}Fe$). Hexagons (diamonds) correspond to L (0.6L). The lines
are only to guide the eye.}\label{fig4}
\end{figure}
\par
In figure 3, we show the percentage deviation $\triangle EVF$ (\%)
of the calculated EVF over experimental data with $\triangle EVF
(\%) = \frac{EVF_{theo}-EVF_{expt}}{EVF_{expt}}\times 100$. The
symbols have same meaning as in figure 2. In the case of IBUU
calculations, percentage deviation for central collisions is about
28\%, whereas for peripheral collisions it is about 10\%. The
average deviation is about 19\% over all colliding geometries. In
the case of IQMD calculations of \cite{liew98}, we see that the
percentage deviation  $\triangle EVF$ (\%) is about 55\% at all
geometries, i.e. the calculated EVF are consistently higher
compared to the data. As mentioned previously and in
\cite{hart98}, in IQMD the system-dependent width of the Gaussian
is used to achieve the maximum stability of the nucleus. It has
also been shown in \cite{hart98,klak93} that the collective flow
and EVF depend strongly on the interaction range of the particle.
The higher the interaction range, the smaller is the collective
flow and therefore larger is the EVF. In IQMD calculations of
\cite{liew98}, the interaction range was taken to be 2 $fm^{2}$
which in part could have led to the reduction of flow and
enhancement of EVF consistently at all colliding geometries. For
our calculations, we use the interaction range 0.6L. We find good
agreement with the experimental data at all colliding geometries.
It is worth mentioning here that the treatment of various
potential terms such as Yukawa, Coulomb and MDI is quite similar
in our IQMD model and the IQMD model of Chen \emph{et al} \cite
{liew98}. Although in our model the range of Yukawa force is 0.4
fm as compared to 1.2 fm in the IQMD model of Chen, it has been
shown in \cite {hart98} that the Yukawa forces have insignificant
effects on collective flow. The treatment of the asymmetry term is
also similar in both the models with C = 32 MeV. Moreover, Pauli
blocking is also treated similarly in both the models. To further
strengthen our point, we therefore took the interaction range L in
our IQMD model and found a huge enhancement in the calculated EVF
making our calculations close to that of \cite{liew98} as shown in
figure 4. It should be noted here that although both $^{58}Fe$ and
$^{58}Ni$ nuclei (see figure 1 and text) show quite similar
stability for Gaussian widths 0.6L and L, the EVF values for the
two different choices of Gaussian width are quite different. This
also indicates that the choice of Gaussian width affects the
collective flow and EVF quite strongly which is in agreement with
\cite{hart98,klak93}.
\par
In Summary, using the IQMD model, we have studied the isospin
effects on the disappearance of flow for the reactions $^{58}Ni$ +
$^{58}Ni$ and $^{58}Fe$ + $^{58}Fe$ (for which data are available)
at all colliding geometries. We have found good agreement of our
calculations with the data at all colliding geometries. Our
calculations explain the data within 5\% (10\%) at central
(peripheral) collisions. We also find that the EVF is affected
strongly by the choice of Gaussian width. We are also able to
explain, in part, the large deviation of the results of
\cite{liew98} from data by the choice of a large width of the
Gaussian.
\par
\textbf{Acknowledgments}\\
 One of the authors, ADS would like to thank
Indian Physics Association for giving national award for Ph.D
thesis in 2006. This work is supported by Indo-French project no.
4104-1.

\end{document}